\documentclass[letterpaper]{jpsj-suppl}
\usepackage{txfonts}

\usepackage{floatrow}

\title{Combined Nucleosynthetic Yields of Multiple First Stars}

\author{Conrad \textsc{Chan}$^{1,2}$ and Alexander \textsc{Heger}$^{1,2,3,4}$}

\inst{%
$^1$Monash University, School of Physics and Astronomy, Victoria 3800, Australia\\
$^2$Joint Institute for Nuclear Astrophysics, 1 Cyclotron
  Laboratory, National Superconducting Cyclotron Laboratory, Michigan
  State University, East Lansing, MI 48824-1321, U.S.A.\\
$^3$University of Minnesota, School of Physics and Astronomy,
  Minneapolis, MN 55455, USA\\
$^4$Shanghai Jiao-Tong University, Department of Physics and
  Astronomy, Shanghai 200240, P.~R.~China}

\email{conrad.chan@monash.edu}

\recdate{September 10, 2016}

\abst{Modern numerical simulations of the formation of the first stars
  predict that the first stars formed in multiples.  In those cases,
  the chemical yields of multiple supernova explosions may have
  contributed to the formation of a next generation star.  We match the
  chemical abundances of the oldest observed stars in the universe to
  a database of theoretical supernova models, to show that it is
  likely that the first stars formed from the ashes of two or more
  progenitors.}

\kword{Pop III stars, nucleosynthesis, supernovae}

\begin{document}
\maketitle

\section{Introduction}

The trace amounts of heavy elements found in observed ultra metal-poor
stars most likely originated from the ashes of the first stars.  The
spectroscopically determined abundance patterns of these metal poor
stars can be matched to the yields calculated from supernova models,
in order to infer the properties of the first stars and their
explosion mechanisms \cite{UN03, HW10}. We have matched the patterns of two metal poor stars in our galactic halo.  Since
the surface layers of these stars have undergone minimal processing of heavier
elements themselves \cite{Wei+04, Pic+04}, they are ideal carriers of the abundance patterns
of their progenitors. Their low abundances also makes it less likely
that they would have been polluted by several independent sources from
unrelated sites.  We assume that the observational uncertainties as provided by spectroscopic abundance determination are accurate and find that individual models do not fit within error bars. Thus we consider the scenario that more than one
progenitor contributed to each observed star, by matching the
summation of multiple yields to observations, with the goal to find
the best fitting combination of models.

\subsection{Model database}

We use a set of Pop III supernova model yields \cite{HW10} calculated using the
stellar hydrodynamics code KEPLER \cite{WZW78,Rau+02}.  This database
has a fine resolution in progenitor mass, explosion energy, and mixing
parametrization.  Since the explosion mechanisms of core collapse
supernovae are not fully understood \cite{Mue16}, we assume that the solution space
is contained within the extensive range of parameters of the model database.  Comparing
these yields to observations provides an approximate starting point to
determining which explosions actually occur.

\subsection{Modelling the combined chemical yields of multiple
  first stars}

We consider the scenario where the entirety of the material ejected by
multiple supernovae mix uniformly such that the chemical composition
of the next generation of stars is represented proportionally by the
mass of the material ejected by each explosion, only diluted by pristine
Big Bang material.  In addition to being a physically motivated assumption,
this also significantly removes the computational cost of determining
the best-fitting ratio, for which physically reasonable solutions are
not guaranteed.

\section{Finding the best fit using evolutionary algorithms}

Evolutionary algorithms are an optimisation method inspired by natural
selection that is able to find approximate solutions, useful when a
complete search for the exact solution is computationally
expensive. Our implementation of the search algorithm,
\textsc{StarFit} (available at \texttt{http://starfit.org}), defines a
population of solutions as combinations of multiple models, and
evolves these solutions through multiple generations until a sufficiently good solution is found. Similar to a genetic algorithm, we apply a crossover operator to solutions in order to combine their information, and a mutation operator in order to fully explore the solution space. At each generation, a selection process takes place to determine the solutions that survive onto the next generation. This selection is weighted towards high-quality fits, but all solutions have a likelihood of being selected in order to preserve diversity. We have tuned the parameters of the search algorithm to efficiently solve our problem. Typically, \textsc{StarFit} is able to find a high quality fit within five minutes on a single thread, rather than the two hours required for a full search.

\begin{figure}[h]
\floatbox[{\capbeside\thisfloatsetup{capbesideposition={right,center},capbesidewidth=0.4\textwidth}}]{figure}[\FBwidth]
{\caption{Graphical representation of the genetic operators. Crossover (left): information from two `parent' solutions (pink, blue) are combined to produce `offspring'. Mutation (right): part of a solution (blue) is randomly changed in order to introduce new information (orange).}}
{
\includegraphics[width = 0.25\textwidth]{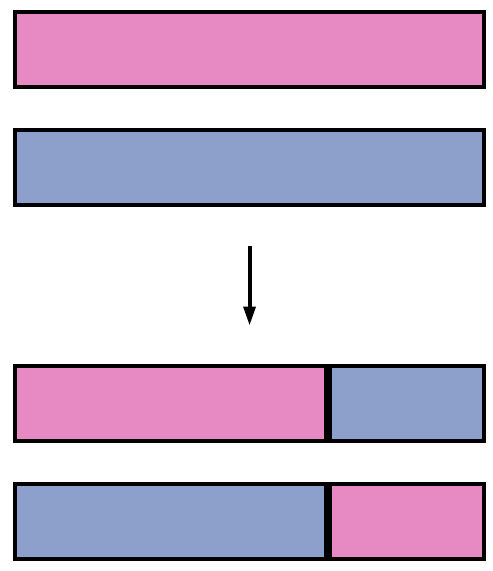}
\qquad
\includegraphics[width = 0.25\textwidth]{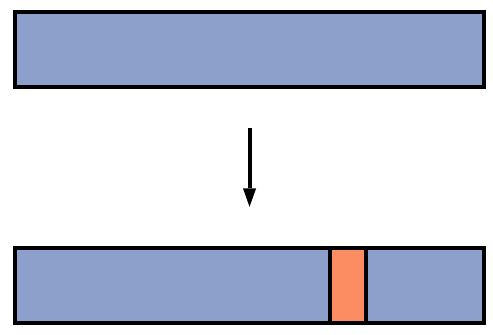}
}
\end{figure}

\section{Results and Conclusions}

We find that for the most iron-poor star, SM0313-6708 \cite{Kel+14}, the yield from a $60\,\mathrm{M}_\odot$ model alone is not sufficient to explain the Li observation, for which the Li-rich $40\,\mathrm{M}_\odot$ yield supplements. This Li is synthesised by the $\nu$-process in the
$40\,\mathrm{M}_\odot$ star which has a more compact H-rich envelope \cite{HW10}.  We also find a small improvement for other elements, which is important given the small uncertainties that are achieved by observations.
Another case that demonstrates the benefit of binary star matching is the iron-poor star HE1327-2327 \cite{Fre+08}, where the CNO-rich yield of a  $75\,\mathrm{M}_\odot$ progenitor supplements a $10.6\,\mathrm{M}_\odot$ yield.

A
possible explanation is the progenitors may have been wide
non-interacting binary stars, which is in agreement with our current
understanding that fragmentation during primordial star formation can
create multiple protostars, often binaries \cite{TAO09, SGB10, Gre+11, Cla+11}.  The supernovae of these
binaries would give a viable path to the formation of stellar-mass
binary black holes in the early universe, a possibility recently
confirmed by gravitational wave detections of mergers \cite{LV+16}.  Another
explanation is that the chemical contributions may originate from
multiple stars in a cluster. Note, however, that our method
  cannot distinguish between true (gravitationally bound) binaries or
  multiple stars and an independent group of these same stars, and also that combined yield scenario is just one possibility.

\begin{figure}[tbh]
\centering
\includegraphics[width=0.49\textwidth]{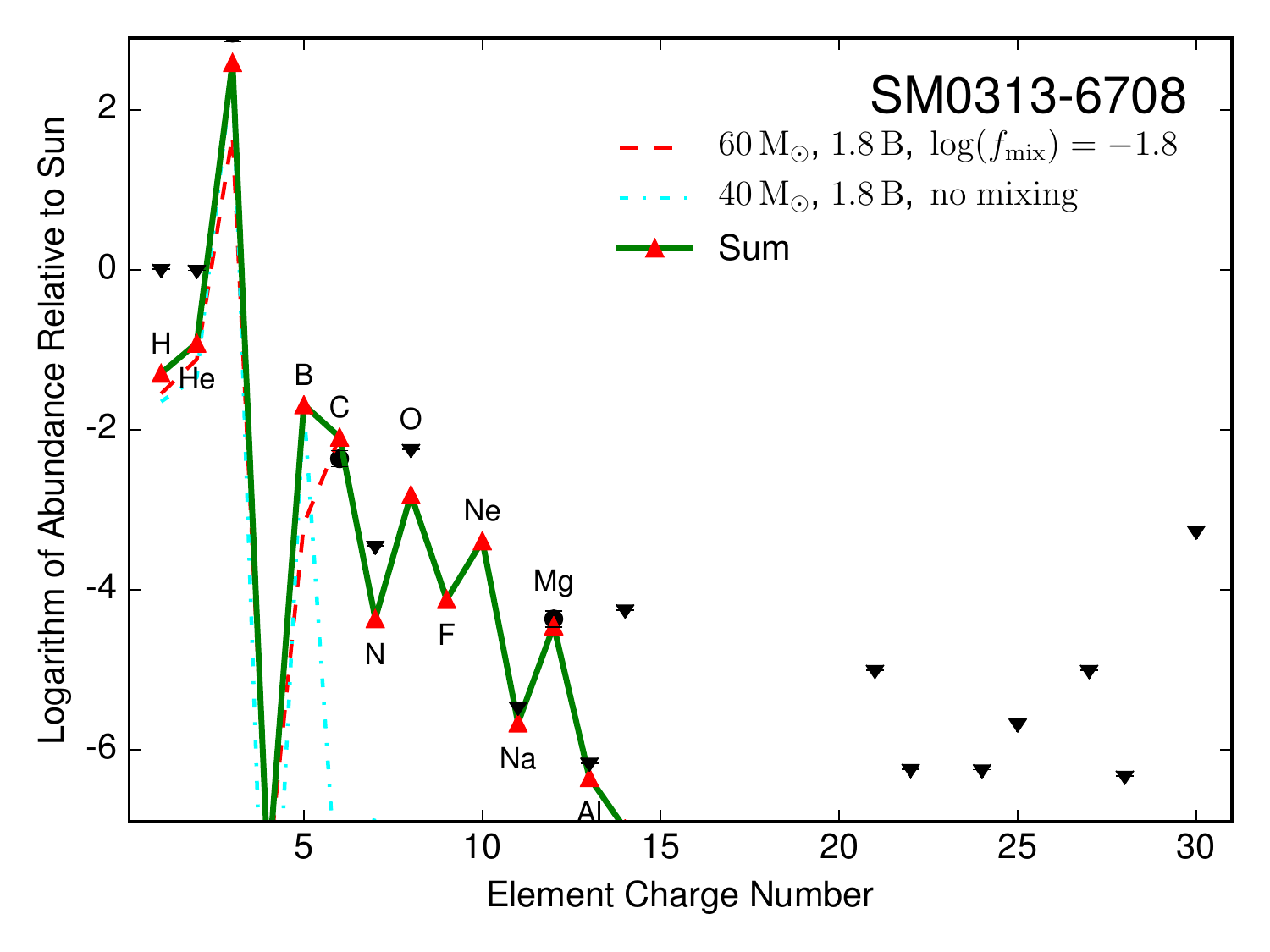}
\includegraphics[width=0.49\textwidth]{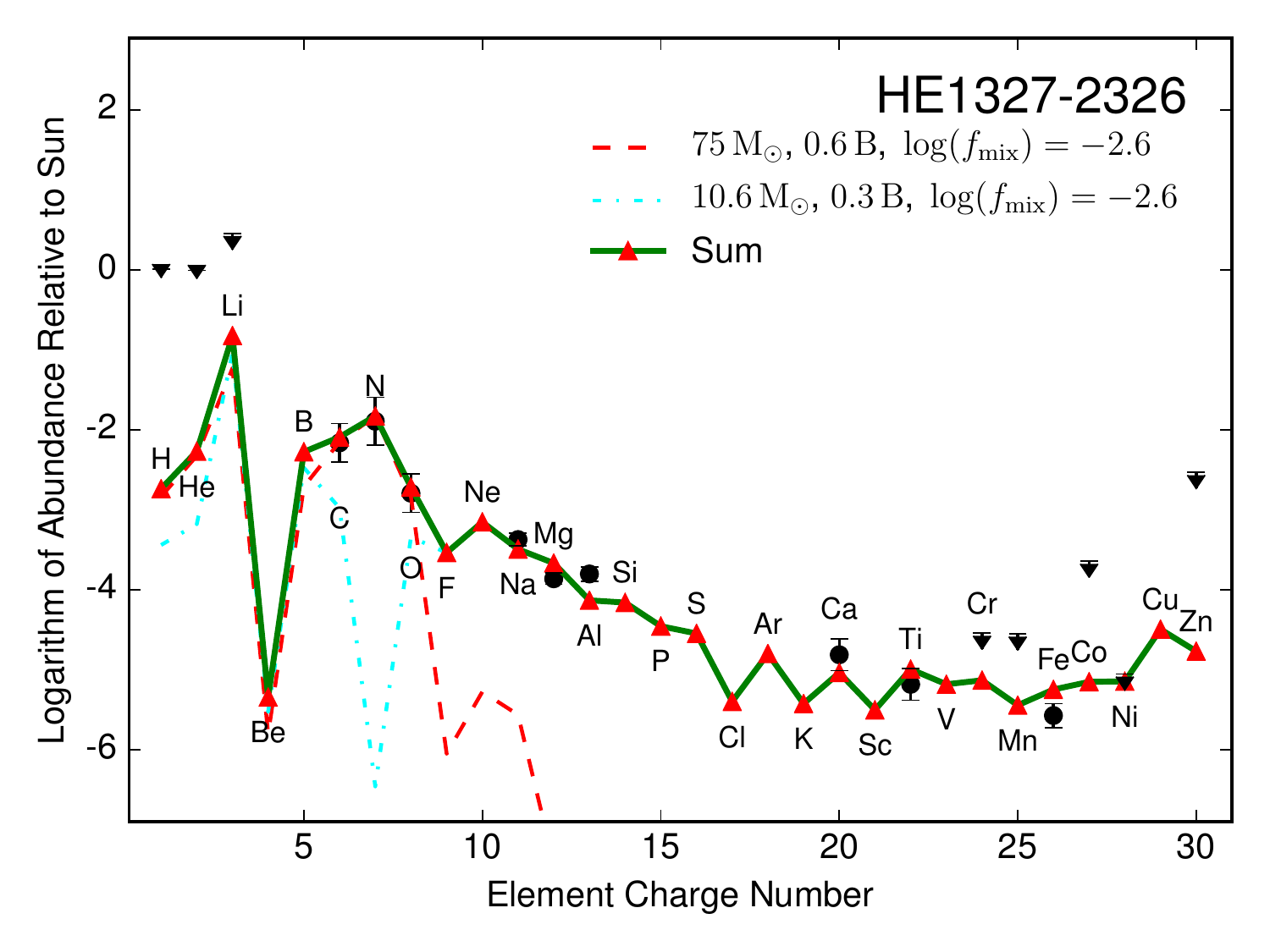}
\caption{Left: The abundance pattern of SM0313-6708
  can be well represented by the combined abundance yields of
  $40\,\mathrm{M}_\odot$ and $60\,\mathrm{M}_\odot$ progenitors. Right: The abundance of HE1327-2326 can be well represented by the combined abundance yields of
  $10.6\,\mathrm{M}_\odot$ and $75\,\mathrm{M}_\odot$ progenitors.}
\label{fig:keller_fit}
\end{figure}

The apparent improvement of fit may be due to an
underestimation of the error bars by observers in the measurement of
spectra, or due to systematic problems in the yields from the
explosion models. The results may be affected by uncertainties in stellar modelling, such as in mixing physics and entrainment, as well as uncertainties in the spectroscopically determined abundances due to NLTE and 3D effects. By considering only non-rotating stellar models, the
database may not cover the necessary parameter space, or the database
may be too coarsely spaced.  The main caveat to our method, however, is
that in general, the model database contains a wide parameterization of mixing and
explosion energy values, many of which are unlikely to be
realistic.  Thus any statistical claims we may make on the uniqueness
of our solutions are at best approximate. Nevertheless, the best
matches found to the database shown here appear to be physically reasonable, and provide an insightful starting point
towards constraining the explosion properties. Upcoming
multi-dimensional simulations of the explosion mechanism and
mixing-fallback with initial conditions motivated by these abundance
matches may confirm whether or not these scenarios are indeed viable.

\end{document}